\documentclass[11pt]{article}

\usepackage[x11names, rgb]{xcolor}

\definecolor{red}{HTML}{f54b1a}
\definecolor{pink}{HTML}{d19eb1}
\definecolor{orange}{HTML}{d3772e}
\definecolor{yellow}{HTML}{ebe85d}
\definecolor{green}{HTML}{0f6852}
\definecolor{lightblue}{HTML}{01abe9}
\definecolor{darkblue}{HTML}{1b346c}
\definecolor{tan}{HTML}{e5c39e}
\definecolor{darktan}{HTML}{af9e73}
\definecolor{grey}{HTML}{c3ced0}
\definecolor{darkgrey}{HTML}{9dadc4}
\definecolor{black}{HTML}{110d1b}
\definecolor{white}{HTML}{f1f8f1}

\usepackage{hyperref}
\hypersetup{
  colorlinks = true,
  linkcolor = red,
  citecolor = darkblue,
  urlcolor = lightblue
}

\usepackage{graphicx}
\usepackage{amsmath, amsfonts, amssymb}
\usepackage[margin=1in]{geometry}

\def\bb{\boldsymbol{b}}

\def\kb{\boldsymbol{k}}
\def\rb{\boldsymbol{r}}

\def\xb{\boldsymbol{x}}
\def\yb{\boldsymbol{y}}

\def\d{\displaystyle}

\def\RR{\mathbb{R}}

\def\<{\langle} \def\>{\rangle}

\def\bb{\boldsymbol{b}}

\def\kb{\boldsymbol{k}}
\def\rb{\boldsymbol{r}}

\def\xb{\boldsymbol{x}}
\def\yb{\boldsymbol{y}}

\def\d{\displaystyle}

\def\RR{\mathbb{R}}

\def\<{\langle} \def\>{\rangle}

\title{Scaling field-theoretic simulation for multi-component mixtures with neural operators}
\date{\today}

\usepackage{fancyhdr, subcaption, textcomp}

\def\d{\textrm{d}}

\graphicspath{{figures/}}

\usepackage[compress, numbers]{natbib}

\begin{document}

\author{Emmit K. Pert, Clay H. Batton, Sherry Li, Steven Dunne, and Grant M. Rotskoff}
\maketitle

\begin{abstract}
    Multi-component polymer mixtures are ubiquitous in biological self-organization but are notoriously difficult to study computationally.
    Plagued by both slow single molecule relaxation times and slow equilibration within dense mixtures, molecular dynamics simulations are typically infeasible at the spatial scales required to study the stability of mesophase structure. 
    Polymer field theories offer an attractive alternative, but analytical calculations are only tractable for mean-field theories and nearby perturbations, constraints that become especially problematic for fluctuation-induced effects such as coacervation.
    Here, we show that a recently developed technique for obtaining numerical solutions to partial differential equations based on operator learning, \emph{neural operators}, lends itself to a highly scalable training strategy by parallelizing per-species operator maps.
    We illustrate the efficacy of our approach on six-component mixtures with randomly selected compositions and that it significantly outperforms the state-of-the-art pseudospectral integrators for field-theoretic simulations, especially as polymer lengths become long. 
\end{abstract}

\section{Introduction}

Field-theoretic simulation (FTS) of a polymer field theory is a versatile~\cite{delaney_theory_2017, fredrickson_coherent_2018}, computationally efficient~\cite{villet_efficient_2014}, and accurate~\cite{mccarty_complete_2019} method for predicting the mesostructure of dense polymer blends. 
This simulation technique scales to systems larger than those in particle-based molecular dynamics by employing a highly coarse-grained representation of the constituent polymers.
The acceleration is achieved by representing each polymer as a free Gaussian chain and subsequently numerically solving for the density in an inhomogeneous chemical potential, representing each polymer species involved not by particles, but by their density fields. 
This approach avoids explicitly computing quadratically scaling pairwise particle-particle interactions and sidesteps the slow relaxation dynamics inherent to long polymers. 
Despite the computational advantages of FTS, multi-component polymer mixtures remain difficult to simulate efficiently because each polymer species must be propagated separately, requiring expensive and repetitive numerical solutions of the ``modified diffusion equation'' (MDE), from which the polymer density fields can be computed. 
Of course, complex mixtures are emblematic of some of the most important open questions in biophysics, as we still seek design principles to explain the self-organization and selectivity of membrane-less organelles \cite{shin_liquid_2017, aumiller_rna-based_2016} and liquid-liquid phase separation, more generally~\cite{brangwynne_germline_2009, najafi_liquidliquid_2021}.

Decades of work has been devoted to improving the numerical algorithms for solving the MDE ~\cite{villet_efficient_2014, ranjan_linear_2008, rasmussen_improved_2002} because it is the most computationally costly part of polymer FTS.
Because this partial differential equation is solved repeatedly with similar chemical potential fields, it is a natural target for acceleration using machine learning. 
In this work, we show that an appropriate formulation of this problem using neural operators~\cite{kovachki_neural_2023} takes advantage of the repetitive nature of this computation, allowing a substantive acceleration of simulation times for complex polymer mixtures. 
Machine learning has, of course, already been explored as a means of accelerating field-theoretic simulations (FTS) with generative adversarial networks \cite{xuan_machine_2023}. However, this approach does not utilize the specific structure of the field-theoretic sampling formalism.   

Our approach to accelerating FTS with machine learning builds on recent work developing scalable strategies to obtain numerical solutions to partial differential equations (PDEs).
The observation that neural networks can serve as powerful ans\"atze for the solutions of PDEs has led to a family of algorithms collectively termed physics-informed neural networks (PINNs)~\cite{raissi_physics-informed_2019, cai_physics-informed_2021,long_pde-net_2019}.
PINNs directly represent the solution of a PDE using a neural network ansatz, and typically rely on a variational formulation of the underlying equation.
While PINNs have shown tremendous utility for a variety of problems~\cite{mao_physics-informed_2020, eivazi_physics-informed_2022,misyris_physics-informed_2020}, the ability of the models to generalize beyond their direct training data continues to be debated~\cite{krishnapriyan_characterizing_2021}.
An alternative approach that focuses on building a mapping purely in the function space, i.e., a \emph{neural operator}, has shown tremendous promise for a variety of problems in physics and beyond~\cite{kovachki_neural_2023, azizzadenesheli_neural_2024, rosofsky_applications_2023}.
Throughout this work, we assess the efficacy of the neural operator approach for field-theoretic simulation, carefully comparing it with the explicit solution of the MDE using traditional operator splitting methods~\cite{duchs_multi-species_2014, pert_coacervation_2024}. 

Ensuring that symmetries and invariances are conserved is a generic challenge that arises when using neural network ans\"{a}tze in computational workflows for physical systems~\cite{thomas_tensor_2018}.
In the neural operator formalism, this issue can be addressed directly by defining an operator that carries out the mapping in Fourier space, akin to the pseudospectral algorithms already widely used to propagate the MDE. 
Indeed, the Fourier Neural Operator (FNO), which we describe in detail below, is well-adapted to the problem of solving the MDE in a field-theoretic simulation workflow, for example, natively satisfying periodic boundary conditions. 
FNOs also have the unique advantage of being invariant to grid density and grid size, making them a natural match for our model, which is also grid invariant.
We show that FNOs accurately represent the ground truth solution of the MDE at significantly lower computational expense.
We give a short introduction to field-theoretic simulation in Sec.~\ref{sec:fts}, and subsequently introduce the basics of neural operators in Sec.~\ref{sec:neuralops}.
We illustrate both the successes and apparent failure modes of this approach in Sec.~\ref{sec:results} and show evidence that suggests that large variations in local density weakly predict error accumulation. 
Overall, the neural operator approach yields strong agreement across all conditions for non-trivial multi-component mixtures, potentially expanding the set of systems for which FTS is feasible. 
All code and input scripts for this work are available on Github at \url{https://github.com/rotskoff-group/polycomp.git}.

\section{Numerical Approach to Field-theoretic Simulation for Polymer Systems}\label{sec:fts}

Statistical field theories have been used to study self-organization in dense polymer melts for decades \cite{fredrickson_field-theoretic_2002, vigil_phase_2023}, both using analytical perturbation theories and numerical simulation. 
While early polymer field-theoretic simulations were mostly limited to single-component systems with simple solvents~\cite{ceniceros_numerical_2004}, the scope of applicability continues to expand to ever more realistic models, including models of biopolymers~\cite{nguyen_molecularly_2023} and industrial synthetic polymers~\cite{fredrickson_coherent_2018}.
To simplify the presentation of both the theory and the numerical results in the present work, we restrict the discussion to uncharged linear polymers, albeit with nontrivial monomer sequences. 
To demonstrate the utility of neural operators in the context of field-theoretic simulation, we focus on mixtures of multiple polymers. 

The field-theoretic representation of a mixture of polymers relies on a mapping between each species' density and its corresponding spatially inhomogeneous chemical potential field.
To fix notation, we walk through the standard development of a field-theoretic representation of a system of free Gaussian chains with repulsive interactions. 
Consider a system of $P$ uncharged polymer types, with $n_j$ polymers per type and $N_j$ monomers each, where $j=1,\dots,P$; here a reference polymer length is denoted by $N$ monomers, and the total number of monomer types is $M$. 
We assume that the polymer is a continuous free Gaussian chain and the bonded interactions along a chain of monomers are given by, 
\begin{equation}
    U_{\rm b}[\rb(s)] = \frac{3\beta^{-1}}{2 b^2} \int_0^{N_j} \left|\frac{\d \rb}{\d s}\right|^2 \d s,
\end{equation}
where $b$ is an effective bond length and $\beta = \frac{1}{k_{\rm B}T}$.
We rescale all lengths relative to the reference $N$ and the bond stretching parameter $b$, leading to an effective radius of gyration $R_{\rm g} = b/\sqrt{N/6}$. 
That is, all coordinates are rescaled such that $\xb$ in $V$ becomes $\rb := \xb / R_{\rm g}$.
The non-bonded interactions are purely repulsive and are specified by the interaction energy
\begin{equation}
U_{\rm nb}(r) = \frac{u_0}{2 \pi a^2} e^{-\frac{r^2}{4a^2}},
\end{equation}
where $a$ has units of length and $u_0$ sets the interaction strength. 
In the Hamiltonian for a homopolymer system consisting only of species $j$ the non-bonded interaction energy is
\begin{equation}
    H_{\rm nb}[\rho_j] = \frac12 \int_V \rho_j(\rb) U_{\rm nb}^{(j,j)}(\|\rb- \rb'\|) \rho_j(\rb')  \d\rb \d\rb' - U_{\rm self}^{(j)},
\end{equation}
where 
\begin{equation}
    \rho_j(\rb) = \sum_{\alpha=1}^{n_j} \sum_{l=1}^{N_j} \delta(\rb - \rb_{\alpha, l})
    \label{eq:deltas}
\end{equation}
and $U_{\rm self}^{(j)} = \tfrac12 n_jN_j u_{\rm nb}(0).$

To make progress, we follow the general computational strategy for polymer field theories~\cite{fredrickson_equilibrium_2006}, i.e., we decouple the non-bonded interactions by introducing auxiliary fields $w_j$ via a Hubbard Stratonovich transformation of the partition function.
First, we view $H_{\rm nb}$ as a functional that acts on a continuous density field $\rho_j$ and write the underlying $\delta$-functions in~\eqref{eq:deltas} as an integral over the inhomogeneous chemical potential field $w_j$. 
This transformation leads to a representation in which the partition function for the system can be written as a functional integral,
\begin{equation}
    \mathcal{Z}(n_j,V,T) = \mathcal{Z}_0 \int \mathcal{D}[\rho_j] \mathcal{D}[w_j] \exp \left( -H[\rho_j, w_j] \right)
    \label{eq:partition}
\end{equation} 
where 
\begin{equation}
    H[\rho_j, w_j] = \frac12 \int_V \rho_j(\rb) U_{\rm nb}^{(j,j)}(\|\rb- \rb'\|) \rho_j(\rb')  \d\rb \d\rb' - i \int_V w_j(\rb) \rho(\rb) \d \rb - n \log Q_j[iw_j],
\end{equation}
and $\mathcal{Z}_0$ is the ideal gas partition function for the polymer degrees of freedom. 
In this expression, $Q_j$ denotes the single-chain partition function associated with the polymer species $j$.

The single-chain partition function can be determined numerically due to the Gaussian chain representation, but the solution depends on the chemical potential field $w_j$.
In particular, $Q_j$ is given by
\begin{equation}
    Q_j = \frac{1}{V} \int_V \d\rb\ q_j \left(\frac{N_j}{N}, \rb \right),
\end{equation}
where $q_j$ solves the modified diffusion equation
\begin{equation}
    \frac{\partial q_j(s, \rb)}{\partial s} = 
    \nabla^2 q_j(s, \rb) - \Gamma * w_j(s, \rb) 
    q_j(s, \rb) ;
    \qquad 
    q_j(0, \rb) = 1.\label{eq:MDE}
\end{equation}
and $w_j(s, \rb)$ is the position-dependent chemical potential field corresponding to the monomer type $j$ parametrically at position $s$ along the curve of the polymer; here, $\Gamma$ is a Gaussian smearing kernel 
\begin{equation}
    \Gamma(\rb) = \left(2 \pi a^2\right)^{-1} \exp \left(-\frac{\rb^T \rb}{2 a^2}\right).
    \label{eq:gauss1}
\end{equation}
Generalizing this field theory to a system of interacting polymers is straightforward, though it does require care around orthogonalizing the fields~\cite{duchs_multi-species_2014}.
Now writing the field theory for the complete set of polymers and evaluating the Gaussian functional integral over $\rho$ in \eqref{eq:partition}, we arrive at a field theory purely in the auxiliary fields $\{w_j\}_{j=1}^n$, which, for notational compactness, we denote as $\{ w_j\}$, 
\begin{equation}
    \mathcal{Z}(\{n_j\},V,T) = \mathcal{Z}_0 \int \mathcal{D}[\{w_j\}] \exp \left( -H[\{w_j\}] \right).
    \label{eq:partition}
\end{equation} 
After integrating out $\{ \rho_j \}$, the Hamilton takes the form, 
\begin{equation}
    H[\{w_i\}] = 
    \sum_{i=1}^M \frac{\gamma_i^2}{2 B_i} 
    \int_V  w_i^2(\rb)\ \d\rb 
    - \sum_{j=1}^{P} n_j \log Q_j[\{w_i\}],
\end{equation}
where $\{B_i\}$ are the rescaled and diagonalized Flory-Huggins parameters, $\{\gamma_i\}$ are Wick rotations that makes all fields real-valued, and $\{w_i\}$ are the chemical potential fields. For further reference, we follow the conventions and notation of~\cite{pert_coacervation_2024} with additional simplifications for uncharged systems. 

The density of each monomer type from the single-chain partition function is given by
\begin{equation}
    \rho_m(\rb) = \frac{C_j}{Q_j}\int_0^{N_j/N} \d s\ q_j(s, \rb) q_j^{\dagger}(s, \rb) \delta (m - m(s)) \label{eq:density},
\end{equation}
where $C_j$ is a reduced concentration, and $q_j^{\dagger}(s, \rb)$ is the adjoint. 
The adjoint is obtained by solving  
\begin{equation}
    -\frac{\partial q_j^{\dagger}(s, \rb)}{\partial s} = 
    \nabla^2 q_j^{\dagger}(s, \rb) - w_j(s, \rb) 
    q_j^{\dagger}(s, \rb) \label{eq:MDEr};
    \qquad 
    q_j^\dagger(\frac{N_j}{N}, \rb) = 1.
\end{equation}
Note that after orthogonalization, the fields $\{ w_j \}$ must be redefined as linear combinations of the chemical potential fields corresponding to the pure monomer types~\cite{pert_coacervation_2024}.
To limit notational complexity, we take $\{ w_j\}$ to be the set of orthogonal fields.

Computing the functional integral $\eqref{eq:partition}$ remains nontrivial even after the Hubbard-Stratonovich transformation.
Although approximate techniques based on saddle point optimization of the model Hamiltonian have been widely used, these mean-field approaches fail to capture important fluctuation-induced effects, such as coacervation~\cite{delaney_theory_2017}.  
While perturbative techniques mitigate these shortcomings to an extent~\cite{sing_recent_2020}, field-theoretic simulations that directly sample the integrand have proved more robust for modeling fluctuation-induced effects~\cite{popov_field-theoretic_2007} because they sample the complete field theory. 
We use the complex Langevin algorithm, which has been widely used for both polymer and quantum field theories~\cite{fredrickson_coherent_2018}.

Equilibrating and sampling a field theory requires iteratively updating the density using the complex Langevin algorithm.
As the name suggests, this algorithm carries out a Langevin-type dynamics on the complex-valued field-theoretic Hamiltonian via, 
\begin{equation}
    \textbf{F} (\boldsymbol{w}(t, \rb)) = 
    \frac{\partial H[\boldsymbol{w}(\rb)]}{\partial \boldsymbol{w}(t, \rb)} 
    = \left(\frac{\boldsymbol{\gamma}^2}{\textbf{B}} \right)
     \odot \boldsymbol w (t, \rb)- 
    \bb^T\boldsymbol{\rho}(t, \rb). 
    \label{eq:cla_force}
\end{equation}
To illustrate the algorithm, we write the scheme using the Euler-Maruyama discretization, and the inhomogeneous chemical potential fields are updated as,
\begin{equation}
    \frac{\partial \boldsymbol{w}(t, \kb)}{\partial t} = - \boldsymbol{\lambda}_w {\textbf{F}}(\boldsymbol{w}(t, \kb)) + \boldsymbol\gamma \odot {\boldsymbol{\eta}}(t, \kb).
    \label{eq:cla}
\end{equation}
Here $\boldsymbol{\eta}$ is a Gaussian random variable with 
\begin{eqnarray}
    \langle\eta_{i\mu}(t, \rb)\rangle &=& 0\\
    \langle\eta_{i\mu}(t, \rb) \eta_{i'\mu}(t', \rb')\rangle &=&
    \frac{2 \lambda_{i\mu} \beta_{i\mu} }{\Delta V } \delta(i-i') \delta(t-t') \delta(\rb-\rb').
\end{eqnarray}
In practice, numerical stiffness in the equation requires that we use more stable numerical schemes, and our numerical experiments integrate~\eqref{eq:cla} in Fourier space with an exponential time differencing method, the exact description of which can be found in \cite{pert_coacervation_2024}. 

Computing the density is by far the most expensive step in conducting field-theoretic simulations. 
As a ``ground-truth'' we solve the MDEs (\eqref{eq:MDE} \& \eqref{eq:MDEr}) pseudospectrally, so the repeated FFTs scale as $O(N \log N)$ in the number of grid points and linearly in the number of polymer segments. 
In our benchmarks, >85\% of total computation time is typically allocated to solving the MDE with a pseudospectral integrator.
We note here that \eqref{eq:density} takes as inputs only $w_j(s, \rb)$.
The field $w_j(s, \rb)$, in turn, only depends on the set of chemical potential fields and information about the polymer architecture.
Because the density is a function of the input fields, we can bypass solving \eqref{eq:MDE} and \eqref{eq:MDEr} explicitly, and instead use a neural operator to directly replace \eqref{eq:density}.
If using the neural operator lowers the computational cost of repeatedly solving the MDE, after accounting for training cost, the use of this technique leads to a substantive boost in efficiency. 
The complex Langevin update step only takes in the current fields and densities, so using a neural operator ansatz to avoid solving the MDE explicitly is an acceptable solution.

\section{Solving the Modified Diffusion Equation with Neural Operators}\label{sec:neuralops}

While a variety of techniques for accelerating the numerical solution of PDEs with machine learning have been developed in recent years~\cite{karniadakis_physics-informed_2021,kovachki_neural_2023}, neural operators are uniquely well-suited to applications in polymer field theory because they can be recombined for distinct mixtures without any additional optimization.  
A neural operator represents the solution of a parametric PDE by learning a mapping between the infinite-dimensional function spaces that represent the input conditions and the output function.
This mapping is learned directly from input-output pairs and is both discretization invariant and highly expressive~\cite{li_fourier_2021}.
Different representations for the neural operator have been proposed, but here, we use the Fourier neural operator (FNO) because of its close connection to pseudospectral integration schemes already widely used for polymer field theories as well as its empirical success on other parabolic PDEs.
The neural operator representation we use here follows directly from~\cite{li_fourier_2021}; in this section, we describe the approach pedagogically for clarity.

\begin{figure}[h!]
    \centering
    \begin{minipage}[c]{0.4\linewidth}
        \centering
        \vspace*{\fill}
        \includegraphics[width=\linewidth]{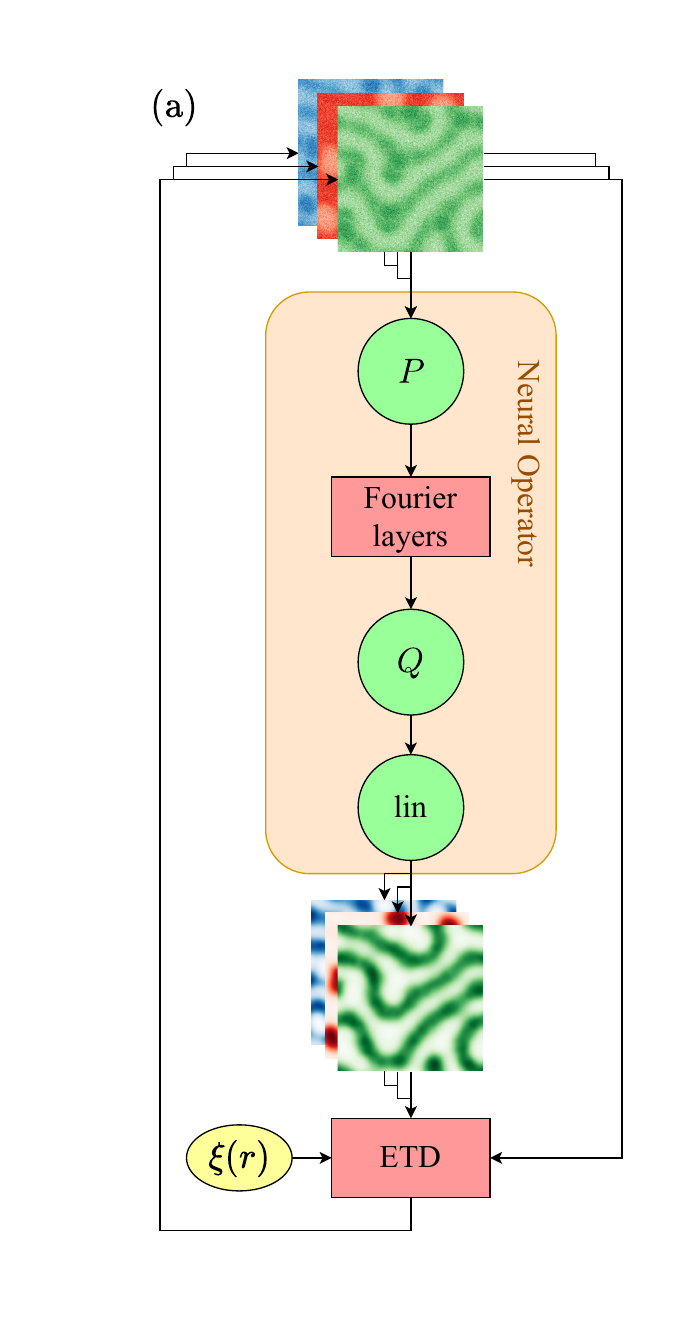}
        \vspace*{\fill}
    \end{minipage}%
    \hfill
    \begin{minipage}[c]{0.55\linewidth}
        \centering
        \includegraphics[width=\linewidth]{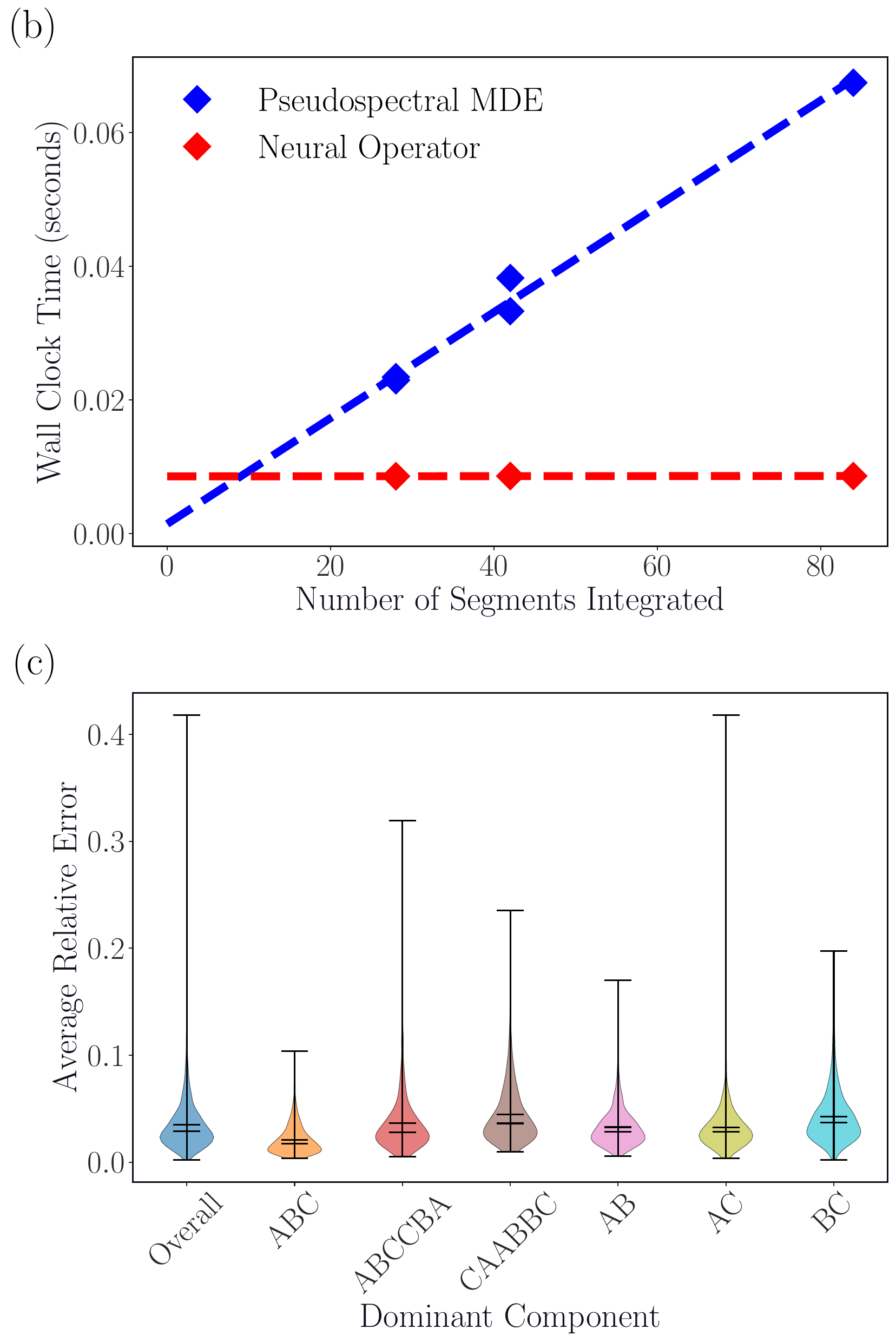}
    \end{minipage}
    \caption{(a) Flowchart diagram illustrating how the neural operator converts chemical potentials (top) and converts them to densities (bottom) that can be used to update the chemical potential through the time integrator. (b) The wall-clock time using the neural operator increases when compared against longer and more complex polymers because these require solving more steps of the MDE. (c) Error for the entire data set and separated by dominant component.}
    \label{fig:main}
\end{figure}

Constructing the operator necessitates defining a learnable (i.e., parameterizable) sequence of functions that can be composed to represent the transformation between the input data and the output solution.
To accelerate the MDE solve, we seek to obtain the species density $\rho$ directly from its associated chemical potential field $w$, effectively bypassing the explicit solve of~\eqref{eq:MDE}.
That is, we seek to learn the map, 
\begin{equation}
    \mathcal{G}^{\dagger}: \mathcal{A}(\RR^d, \RR^{d_a}) \to \mathcal{U}(\RR^d, \RR^{d_u}); \quad \mathcal{G}^{\dagger} w(\rb) \mapsto \rho(\rb),
\end{equation}
which is defined on the separable Banach spaces $\mathcal{A}$ and $\mathcal{U}$; schematically, the application of this operator yields the solution of the PDE from input data specified through a function in $\mathcal{A}$.
The main idea of the neural operator approach is to represent this transformation by learning a sequence of functions $v_t$ in a lifted space with coordinates $\xb$ and composing them to transform the input function.
Pointwise, these transformations are defined recursively as,
\begin{equation}
    v_{t}(\xb) = \sigma \bigl( W_{t-1} v_{t-1}(\xb) + \mathcal{K}_{t-1}[v_{t-1}](\xb) + b_{t-1}(\xb) \bigr),
\end{equation}
where $W\in \RR^{d_{v_{t}}\times d_{v_{t-1}}}$ is a linear weight matrix, $b:\RR^{d_{v_{t-1}}}\to \RR^{d_{v_{t}}}$ is a bias function, $\sigma$ is a nonlinear activation function applied entrywise, and $t$ ranges from $1,\dots,T$ which sets the ``depth'' of the neural operator.
In this expression, a key ingredient is the integral operator,
\begin{equation}
    \left(\mathcal{K}_t[v_t]\right)(\cdot ) = \int_{D_t} \kappa_{\theta}^{(t)}(\cdot,\yb)v_t(\yb) \d \yb,
    \label{eq:intker}
\end{equation}
where $\kappa^{(t)}_{\theta}:\RR^{d_{v_t}}\times D_t \to \RR^{d_{v_{t+1}}}$ is a parametric function playing the role of an integral kernel. 
These transformations operate in a lifted space; that is, we first use a standard feed-forward neural network such as a multilayer perceptron to ``lift'' the input function to produce $v_1$.
Mathematically, we define an operator $\mathcal{P}:\RR^d \to \RR^{d_{v_1}}$ such that 
\begin{math}
    \mathcal{P}:w(\rb) \mapsto v_1(\rb)
\end{math}.
Similarly, to recover the output density function, we project using $\mathcal{Q}:\mathbb{R}^{d_{v_T}} \rightarrow \mathbb{R}^{d}$.
The projector is also represented by a feed-forward neural network.
With the above definitions, we then seek to approximate $\mathcal{G}^{\dagger}$ with the neural operator,
\begin{equation}
    \mathcal{G}_\theta := \mathcal{Q} \circ \sigma \left(W_{T} + \mathcal{K}_{T} + b_{T}\right) \circ ... \circ \sigma \left(W_1 + \mathcal{K}_1 + b_1\right) \circ \mathcal{P}.
\end{equation}
The overall architecture of this approach is illustrated in Fig.~\ref{fig:main} (a).

To build an efficient and flexible transformation using the integral kernel $\mathcal{K}$, it is convenient to work with the transformation in Fourier space, just as is done with pseudospectral integration schemes because it allows computationally efficient evaluation of the integral \eqref{eq:intker}.
Indeed, because the convolution with the kernel $\kappa^t$ can be expressed as a product in Fourier space due to the convolution theorem, the Fourier Neural Operator employs the representation,
\begin{equation}
    v_{t+1}(\xb) = \mathcal{F}^{-1}\left(R_\phi \cdot \mathcal{F}(v_t)\right) (\xb),
\end{equation}
where $\mathcal{F}$ is the Fourier transform and $R_\phi$ is simply a matrix of parameters learned by the model. 
Though this choice of $R_{\phi}$ appears simple, previous benchmarks have established that it provides the most efficacious integral kernel for a variety of PDEs~\cite{li_fourier_2021}. 

Throughout, we assume that $\kappa^{(t)}$ is a periodic function and hence admits an expansion into discrete Fourier modes.
For each transformation $v$ mapping to a space of dimension $d_v$, the dimensions of $R_{\phi}\in \mathbb{C}^{k_{\rm max}\times d_v \times d_v}$ set an effective truncation at mode $k_{\rm max}$, which we choose to balance performance and error.
Hence, the kernel integration can be expressed in Fourier space by the contraction
\begin{equation}
    \left( R_\phi \cdot (\mathcal{F} v_t) \right)_{k,l} = \sum_{j=1}^{d_{v_t}} R_{k,l,j}^{(\phi)} (\mathcal{F} v)_{k,j}, \quad k = 1, \ldots, k_{\text{max}}, \quad l = 1, \ldots, d_{v_{t+1}},
\end{equation}
 where $d_{v_t}$ and $d_{v_{t+1}}$ are the dimensionalities of the lifted layers $v_{t}$ and $v_{t+1}$, respectively. 
 
 Training the Fourier Neural Operator follows the standard workflow supervised learning with gradient-based optimization.
 Given a collection of $n$ ground truth pairs, $\{(w^{(i)}, \rho^{(i)})\}_{i=1}^n$ define an empirical loss function,
 \begin{equation}
    \mathcal{L}(\theta) = \frac{1}{n} \sum_{i=1}^n \| \rho^{(i)} - \mathcal{G}_\theta(w^{(i)}) \|^2,
 \end{equation}
 where $\theta$ denotes the full set of parameters in the FNO model.
 We optimize this objective using stochastic gradient descent-based algorithms, such as adam~\cite{kingma_adam_2017}.
 Losses including higher-order derivative terms such as Sobolev losses can also be used for optimization.
 In our numerical experiments, we employ the $H^1$ Sobolev loss. 
 
 We carry out field-theoretic simulations using the trained FNOs using complex Langevin simulation via~\eqref{eq:cla}. 
 Each update to the fields $\{w_j\}$ then yields a new corresponding density field $\{ \rho_j \}$ which we compute directly via a forward pass using $\mathcal{G}_{\theta}^{(j)}$.
 We emphasize that each species has a distinct neural operator but that these operators are trained only on independent simulations of each species in isolation. 
This fact has important implications for the cost associated with computing, e.g., phase diagrams in which the concentrations of various species are varied.

\section{Results} \label{sec:results}

\begin{figure}[h]
    \centering
    \includegraphics[width=0.99\linewidth]{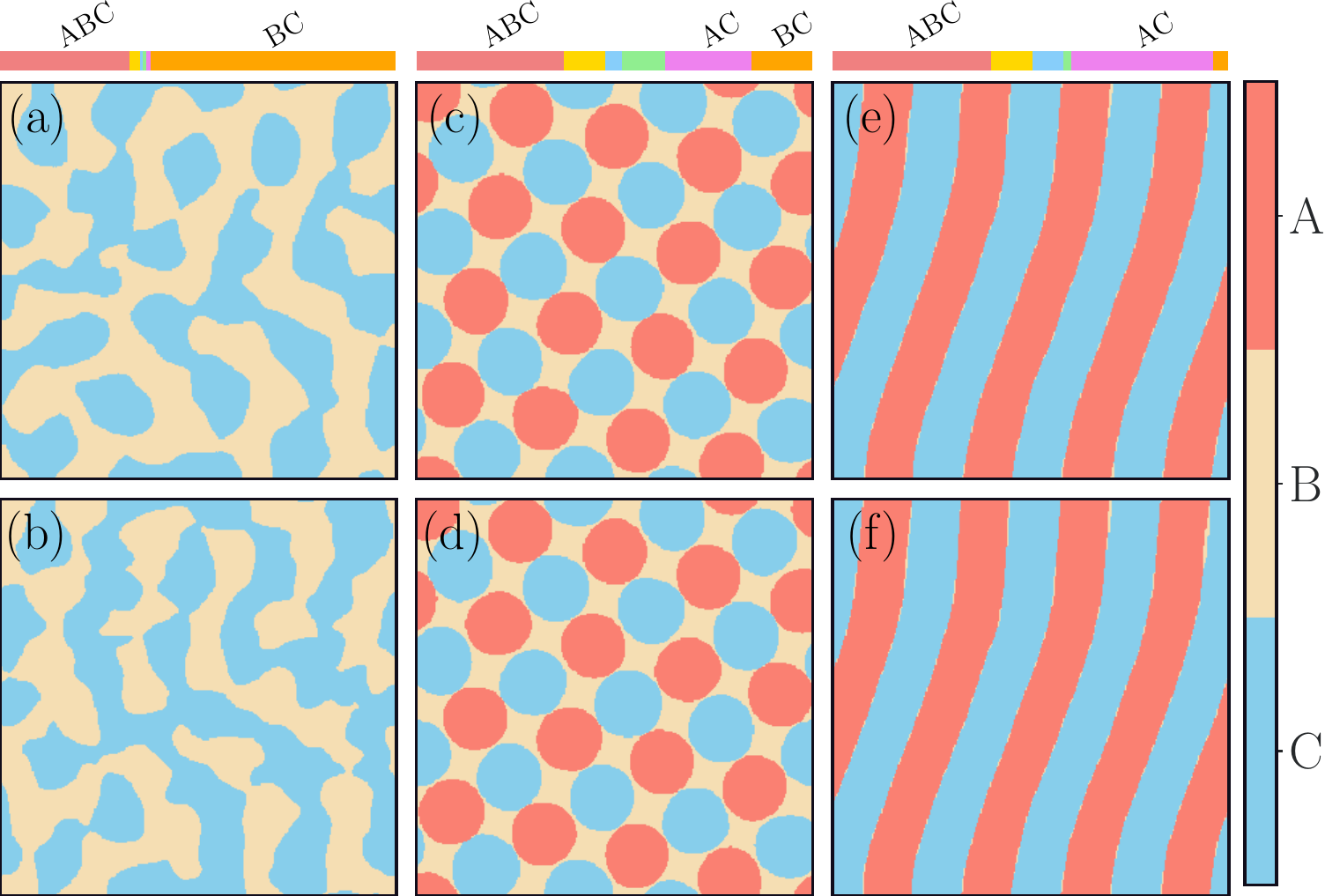}
    \caption{Three of the ten best-performing structures (a,b,c) with their corresponding numerical-solved densities (e,d,f). The average errors are (a) 0.22\%, (b) 0.35\%, (c) 0.38\%, these were selected to show some of the best-performing examples that had varied morphologies. The top color bars represent the system composition, with the highest concentration species labeled explicitly. }
    \label{fig:good_examples}
\end{figure}
\begin{figure}
    \centering
    \includegraphics[width=0.5\linewidth]{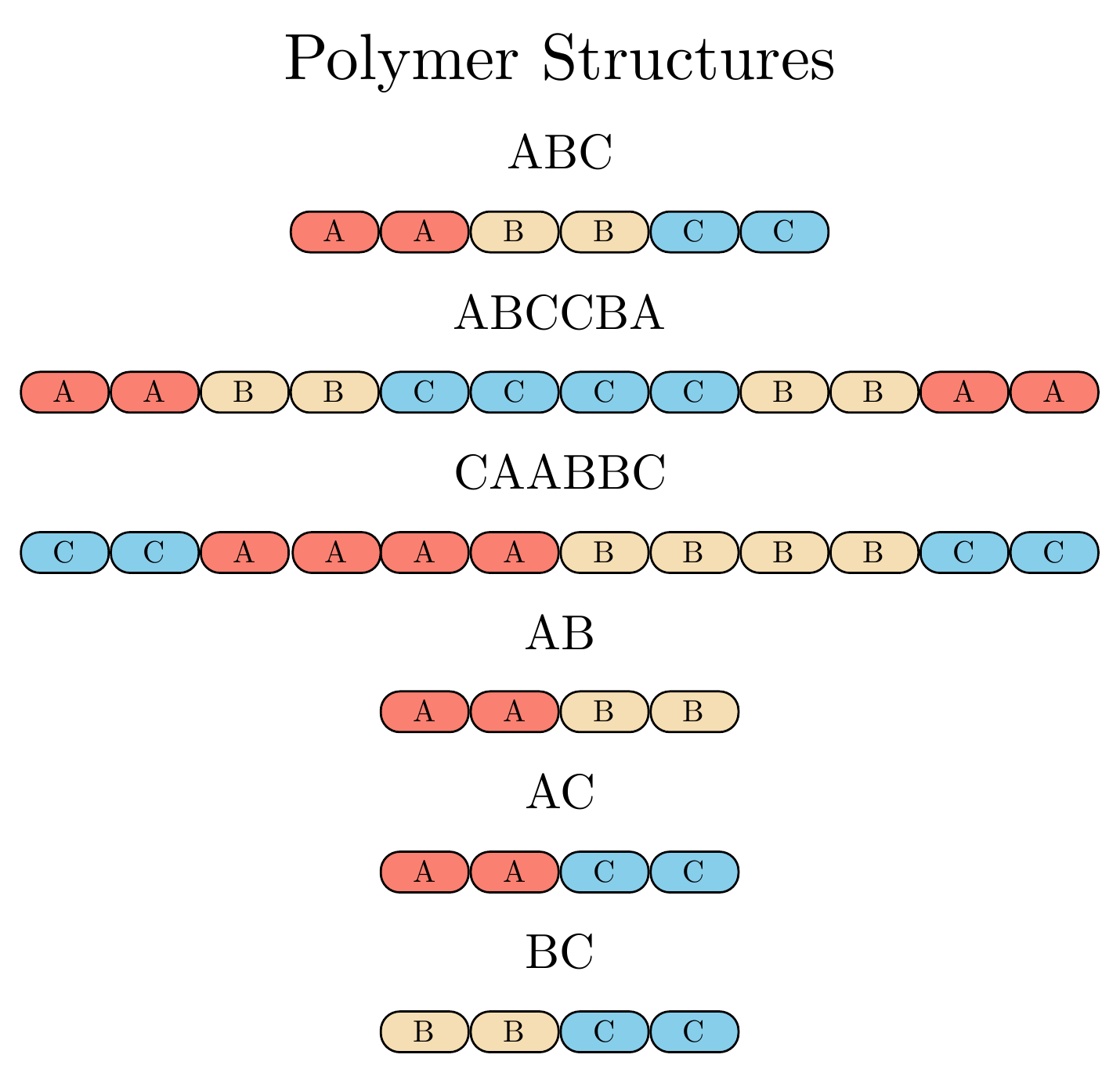}
    \caption{Schematic of the six polymer structures included in the mixed system. Each segment is partitioned into more than two blocks in simulation}
    \label{fig:structures}
\end{figure}

Multi-component polymer mixtures constitute an important, but still largely intractable, problem domain for polymer field-theoretic simulation.
We narrow our focus on this setting and study the performance of the FNO approach on a six-component mixture of block copolymers. 
The six components we consider are polymers composed of three distinct monomer types, denoted $A$, $B$, and $C$. 
Denoting a block of fixed length by a single letter, we simulated mixtures of six block copolymers with structure ABC, ABCCBA, CAABBC, AB, AC, and BC, as shown in Fig.~\ref{fig:structures}. 
We constructed the Flory-Huggins matrix so that all polymers microphase separate in isolation and all interactions would yield a symmetric Flory-Huggins interaction matrix, $B$. 
We set the diagonal $B_{ii} = 130$ and the cross terms $B_{ij} = 151$, because it leads to microphase separation while also robustly maintaining numerical stability. 
To mitigate problem selection bias, we chose all parameters before any training was conducted, and they were not modified thereafter. 
A collection of configurations representative of the morphological diversity of these mixtures which also illustrates the accuracy of the neural operator approach is shown in Fig.~\ref{fig:good_examples}.
In these examples, the neural operator result is qualitatively indistinguishable from solutions obtained via the pseudospectral method, even when integrated over thousands of steps of dynamics.
For computational details of the neural operator and the field-theoretic simulations, see Sec.~\ref{sec:comp_details}.

To compose the training set, we began by running several trajectories of a single polymer type and saving uniform random samples of the chemical potential along these trajectories, with further details in Sec.~\ref{sec:comp_details}. 
For each polymer type, the final training set included all the chemical potentials and the density corresponding to the polymer type of interest conditioned on the given chemical potential, which had to be computed. 
Crucially, the training data did not include structures containing mixed blends of multiple polymers---FNOs for each independent species were trained on simulations conducted with a single polymer type.
We note here that any set of chemical potentials for which the density can be calculated is a valid training set, so there may be more efficient ways to generate training data.
Indeed, we believe that it may be possible to find collections of training configurations that span the set of densities that are physically accessible within a class of systems of interest. 

\begin{figure}[h]
    \centering
    \includegraphics[width=0.95\linewidth]{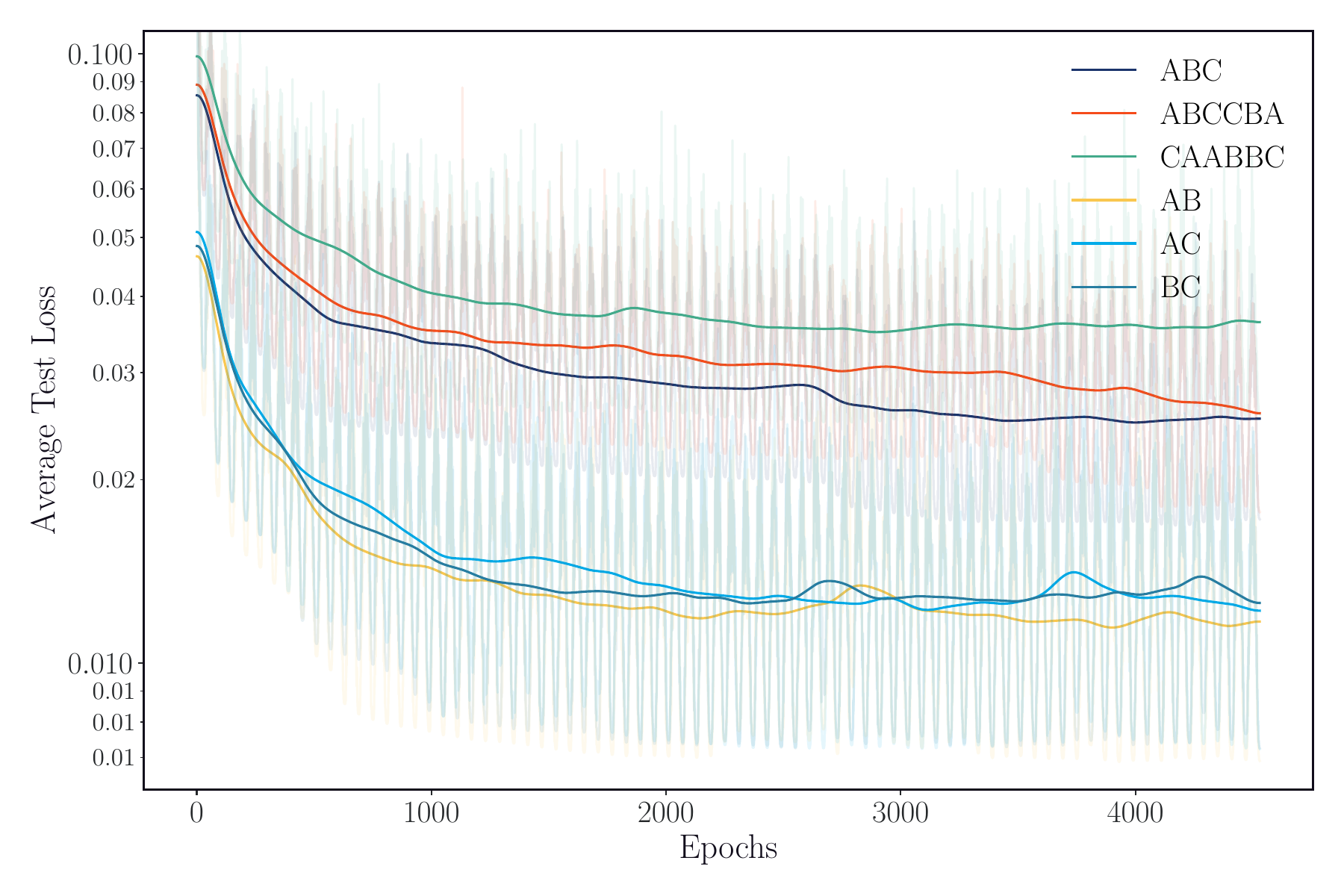}
    \caption{Test loss for each of the 6 species. Exact losses with cosine annealing in partial transparency with moving average in the foreground. The test loss is computed by reserving 12.5\% of the dataset for validation, with the loss being computed by an average of the $H^1$ norm over each species. Note that the diblock loss is lower due to having an absent density that is still present in the average.}
    \label{fig:training}
\end{figure}

In our case, a separate neural operator was trained for each polymer, which then was used to replace the modified diffusion equation as previously described.
The training was conducted with a cosine annealing rate schedule using a Sobolev loss, implemented from the FNO code base~\cite{li_fourier_2021,kovachki_neural_2023}.
The test loss was computed by reserving 12.5\% of the data to evaluate for unseen chemical potentials. 
Training was run for a fixed time and test results were examined to determine whether the data was sufficiently converged.
We have included training curves for all six of the polymers in Fig.~\ref{fig:training}, with the diblocks achieving a lower loss due to having a density of zero in the absent component which is included in the average.

\begin{figure}[h!]
    \centering
    \includegraphics[width=0.99\linewidth]{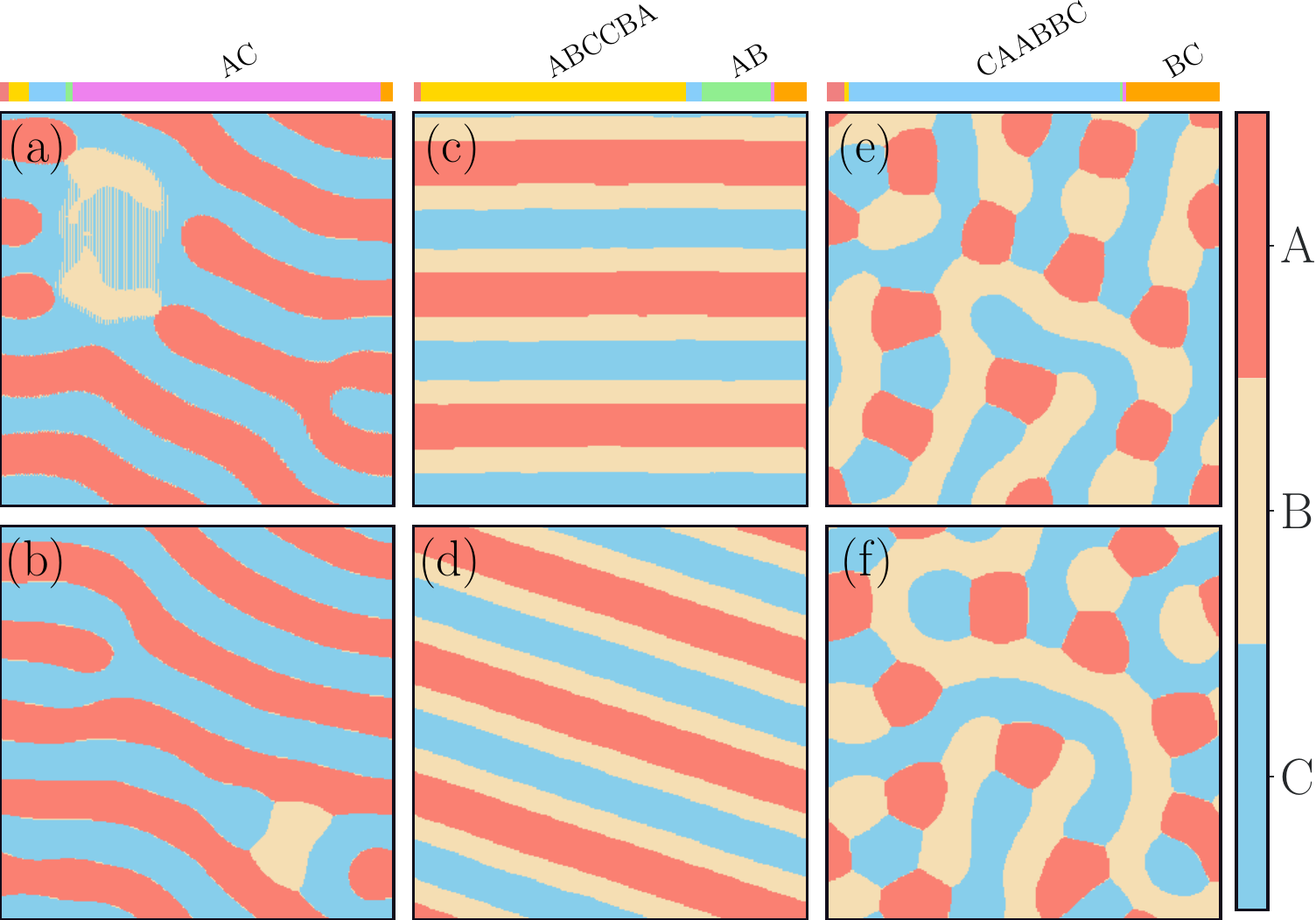}
    \caption{Three of the four worst performing structures (a,c,e) with their corresponding ground-truth pseudospectral MDE densities (b,d,f). The average errors are (a) 41.7\%, (b) 24.2\%, and (c) 22.5\%, these were selected to show the worst performers that had different morphologies.}
    \label{fig:bad_examples}
\end{figure}

To test the model accuracy, we generated 5000 random mixtures of the six polymer types uniformly distributed on the simplex of allowable compositions \cite{devroye_non-uniform_1986} and ran dynamics to equilibrium with the neural operator. 
The error was calculated by computing the density from direct numerical solution of the modified diffusion equation as a ground truth. 
We quantified the relative error 
\begin{equation} \label{eq:error}
    \textrm{err}_{\textrm{rel}} = \frac{1}{3V}\sum_M\int_{V} \frac{|\rho_{N,M}(r) - \rho_{M}(r)|}{\rho(r)_{M}} \d r.
\end{equation}
in the densities of each of the three monomer types and then averaged over the monomer types to obtain the relative error, where $ \rho_{N,M} $ and $ \rho_{M} $ are the monomer densities of type $ M $ obtained from the numerical solution and neural operator, respectively.
The reported errors are obtained by averaging over the last third of the simulations with a stride of 50 time points.
Figure~\ref{fig:main} (c) shows the distribution of $\textrm{err}_{\textrm{rel}}$ across all mixtures, indicating a high degree of accuracy in the density, excepting some rare outliers that sustain large errors. 
While the test losses obtained for each component are variable and, indeed, are highest for the three component polymers, as shown in Fig.~\ref{fig:training}, the contribution to the overall error is not predicted by the relative error in practice. 
In fact, in simulations in which the ABC polymer is the dominant component, the fewest outliers with high error are observed. 
We show morphologies of the mixtures that perform the worst in Fig.~\ref{fig:bad_examples}, where we plot the dominant component in the mixture at each point.

\begin{figure}
    \centering
    \includegraphics[width=0.99\linewidth]{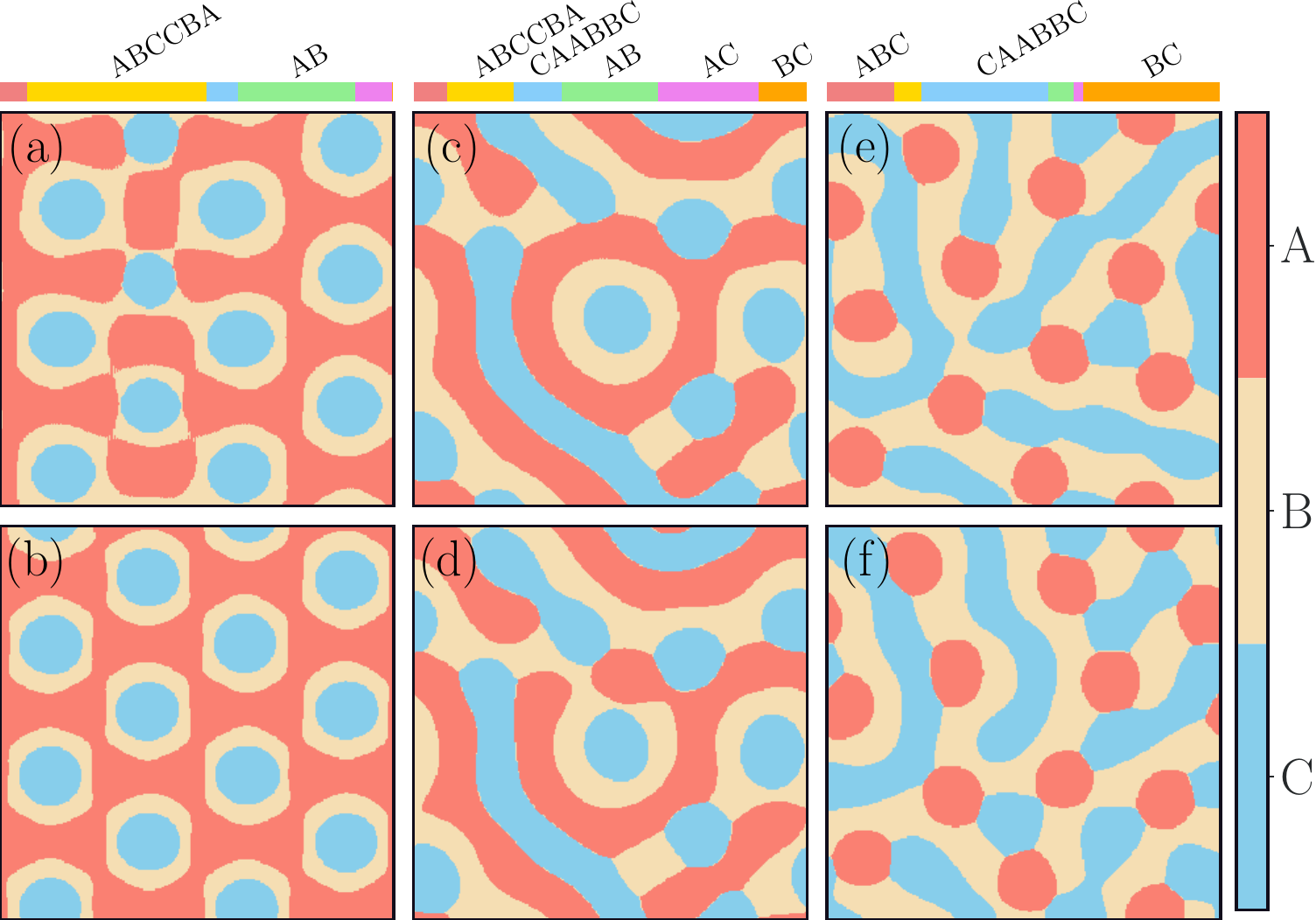}
    \caption{3 randomly selected plots to help illustrate the diversity of morphologies correctly predicted by the model. Errors are (a) 6.8\%, (b) 2.5\%, (c) 7.8\%}
    \label{fig:random_examples}
\end{figure}

The morphology of the samples was highly variable and showed many distinct phases, which we do not exhaustively catalog here. 
We have included a random sample of some of the phases in the supplement to indicate that the model is not limited only to one particular phase, but is robust across a wide variety of structures Fig.~\ref{fig:random_examples}.
Qualitatively, the best-performing examples shown in Fig.~\ref{fig:good_examples} have relatively low complexity compared with a typical or high error sample. 
The anomalously erroneous examples in Fig.~\ref{fig:bad_examples} have one or two species that are highly over-represented or evince significant spatial dislocations. 
Some of these simulations also show visible aliasing in the density field, which represents a failure of the model to predict the density accurately. 
The aliasing generally arises due to insufficient $k_{\rm max}$ when truncating a Fourier basis, which could also be the origin of the error for the FNO.
Indeed, the FNO uses a cutoff frequency, but we found that increasing that parameter beyond the value used here did not lead to stable training. 
Additionally, while it is not visible in the static images, some configurations demonstrate a linear drift instability which can also rise in pseudospectral FTS near the limit of stability. 
More generally, ``out-of-distribution'' generalization for the FNO ansatz appears to be somewhat weak, though this can be mitigated with inexpensive single-polymer training over larger chemical potential ranges.
The single worst example, Fig.~\ref{fig:bad_examples}(a), is one in which the model aggregates a single species into a highly localized region of space producing densities not seen during training, yielding a large error compared to the typical examples. 

We realize a significant computational advantage for the neural operator formalism when amortizing the cost of training, as illustrated in Fig.~\ref{fig:main} (b).
The cost of running inference does not depend on the length or complexity of the polymer structure, unlike pseudospectral methods, which have a linear dependence on polymer length. 
We see that the wall-clock time required to run eight segments of pseudospectral integration coincides with neural operator evaluation for polymers of arbitrary length. 
Because polymers that are not fully symmetric have to be integrated in both directions, this is equivalent to a polymer with four integration blocks if it is asymmetric, which is far shorter than most polymers of interest. 
While this analysis amortizes training cost, in systems where we are interested in many different compositions, this fixed cost is insignificant compared to repeated inference.

\section{Conclusions}

We have demonstrated that Fourier neural operators provide an efficient, composable alternative to numerically solving the modified diffusion equation. 
This approach yields substantial computational advantages, particularly for systems consisting of mixtures of polymers, where the training cost can be effectively amortized due to the high degree of transferability between systems. 
This operator approach robustly represents the full spectrum of different mixtures and efficiently scales to large systems. 
The agreement between the learned transformation and the numerical one is, in large part, excellent, though conditions that push the dynamics far from the training domain can lead to anomalous poor performance. 

An intuitive scaling analysis demonstrates that the advantage of our approach arises for long and complex polymers, which require dense discretization for numerical stability and accuracy. 
The neural operator MDE solve is independent of the polymer length, whereas the pseudo-spectral integrator scales linearly with the length. 
In addition to the computational advantages for long polymers, the model is highly transferable. 
Because the propagator for each polymer is only a function of the chemical potential and is thus independent of other species in the mixture, neural operators are particularly well-suited to multi-component mixtures.
Because we independently train one neural operator per polymer type, the representations are reusable in arbitrary mixtures. 
We observe a high degree of \emph{transferability}, i.e., we see that this approach generalizes well to new mixtures, massively reducing the cost of studying mixtures as a function of concentration.

Learned representations of function space mappings carry with them an inherent error, just like other neural representations. 
While we find strong agreement throughout, we characterize two rare failure modes: out-of-domain error associated with strong concentration of density of a single species and aliasing errors related to high-frequency Fourier components. 
The out-of-domain error can be corrected by refining the training data generation procedure, either by active learning or some other method of generating expressive data inputs. 
The aliasing issue appears most severe in cases where we are in unusual configurations but appears to be linked to the underlying operator basis.
Additionally, it is straightforward to detect deviations from exact numerical results by using intermittent evaluations of the density per \eqref{eq:density}, allowing increased trust in the results at minimal additional expense. 

We expect that with limited issues, successful implementation with relatively little tuning, and substantial performance improvements, the methodology should be widely applicable to many polymer systems.
Applications to biological systems such as membraneless organelles appear promising because this approach enables modeling protein aggregation while mitigating the cost of simulating many different biopolymer species. 
This method for efficiently calculating polymer densities may enable a new computational paradigm for studying these systems at scale. 

A number of extensions of the framework laid out here remain important for future applications. 
Generalizing the approach to charged systems will be an important aspect of capturing coacervation. 
Scaling data generation will also be crucial because the training cost requires explicit solutions to the MDE: an active learning strategy to select the chemical potential fields for exact solution could bring this cost down substantially. 
Generalizing the model to be ``transferable'' across polymer species would compound this acceleration.

\bibliographystyle{plain}
\bibliography{references, ref_emmit}

\section{Supplementary Information}

\subsection{Computational Details} \label{sec:comp_details}

The Fourier Neural Operator was used as implemented by the FNO package \cite{li_fourier_2021}. 
The TFNO model was built in 2d with $k_{\rm{max}}$ of (32,32), 32 hidden channels, 64 projection channels, Tucker factorization, and rank of 0.42. 
The training was conducted with a learning rate of 0.002, weight decay of 0.0001 and cosine annealing with a $\rm{T}_{\rm{max}}$ of 30 for 4530 epochs - enough time to show test loss (H1 Sobolev norm) leveling off. We did not conduct extensive hyperparameter tuning, so we expect that there may be some gains to be made in improving these parameters.

For all field-theoretic simulations, we used a relaxation rate of 2.25 and a temperature of 0.001. All same-species FH parameters were 130 and all different species interactions were 151 and grid smearing coefficient $\alpha=0.2$.  The polymer had an integration width of 1/20, indicating that a polymer of length 1 would be broken into 20 integration points. For all polymer species, a single letter in the name corresponds to a block of size 1/3, so for instance the ABC polymer has 3 1/3 blocks - one for each polymer and a total length of 1.  

To generate the training data, we ran 200 simulations of length 3000 (a time selected to allow equilibration) for each polymer, with grids randomly sized between 8x8 and 18x18 and initialized at a homogeneous state. From these trajectories each chemical potential was randomly saved with probability 1/600. These chemical potentials were combined into a single dataset and the corresponding density was computed for each polymer type for every saved chemical potential. 

The actual production simulations were run with a box size of 16x16 with 128x128 grid points for 6000 steps with a relaxation rate of 1.5.
All of the parameters are equivalent for the generation of the training data unless otherwise specified.
Proposed densities for each species were sampled uniformly on the simplex by generating 5 random numbers between 0 and 1, and making the concentration of each species equal to the difference between the numbers when ordered in increasing value.

%
%
%
%

\end{document}